\documentclass[twocolumn,showpacs,preprintnumbers,amsmath,amssymb]{revtex4}
\usepackage{tabularx,graphicx}\begin{document}
\newcommand{\beq}{\begin{equation}}
\newcommand{\eeq}{\end{equation}}
\newcommand{\beqn}{\begin{eqnarray}}
\newcommand{\eeqn}{\end{eqnarray}}
\newcommand{\bmath}{\begin{subequations}}
\newcommand{\emath}{\end{subequations}}
\title{Charge expulsion, charge inhomogeneity and phase separation in dynamic Hubbard models}
\author{J. E. Hirsch }
\address{Department of Physics, University of California, San Diego\\
La Jolla, CA 92093-0319}
 
\date{\today} 
\begin{abstract} 
Dynamic Hubbard models are extensions of the conventional Hubbard model that take into account the fact that atomic orbitals
expand upon double occupancy.  It is shown here that systems described by dynamic Hubbard models have a tendency to
expel negative charge from their interior to the surface, and  to develop charge inhomogeneity and even phase separation in the bulk. 
These effects are associated with lowering of electronic kinetic energy. We propose that these models may explain the    charge inhomogeneity and negatively charged grain boundaries 
observed in cuprate oxides and other materials.
\end{abstract}
\maketitle 

\section{Introduction}

 The conventional single band Hubbard model with Hamiltonian
\beq
H=-\sum_{i,j,\sigma}[t_{ij}c_{i\sigma}^\dagger c_{j\sigma}+h.c.]+U\sum_i n_{i\uparrow}n_{i\downarrow}
\eeq
has been used to describe the physics of many real materials\cite{hub,hub2}. The model
ignores the fact that non-degenerate atomic orbitals are necessarily modified by double electronic occupancy\cite{orbx,hole1,relax,inapp}. To remedy this deficiency a variety of new Hamiltonians have
been proposed and studied that we will generically call `dynamic Hubbard models', that take into account the fact
that {\it orbital expansion} takes place when a non-degenerate atomic orbital is 
doubly occupied\cite{tang,hole2,dynhub,dyn3,dyn5,dyn7,dyn8,dyn11,dyn121,bach1,dyn12}.

The essential physics of real atoms that is described by dynamic Hubbard models but not by  the conventional Hubbard model is shown in Fig. 1: when a second electron (of opposite spin) is added to a non-degenerate orbital, it expands, due to electron-electron repulsion. This has two key consequences at the atomic level. One is, negative charge moves outward.
The second is, the kinetic energy of the electrons is lowered: in an orbital of radial extent  $r$ the electron kinetic energy is of order $\hbar^2/(2m_e r^2)$, with $m_e$ the electron mass.
The kinetic energy is lowered since the expanded orbital has larger radius than the original one. Remarkably, we will find that these properties at the atomic level,
negative charge expulsion and kinetic energy lowering, are also reflected in the
properties of dynamic Hubbard models at the macroscopic level. At the local level, effects described by the dynamic Hubbard model that are not described by the
conventional Hubbard model can be experimentally probed by ultrafast quantum modulation spectroscopy as recently demonstrated\cite{kaiser}.

 \begin{figure}
\resizebox{8.5cm}{!}{\includegraphics[width=7cm]{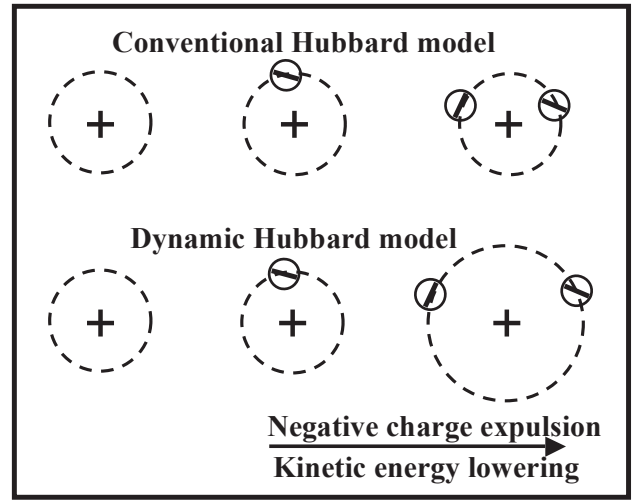}}
  \caption{In the conventional Hubbard model the atomic orbital is not modified by electronic occupancy. In the dynamic Hubbard model
  and in real atoms, addition of
  the second electron causes orbital expansion due to the electron-electron interaction. Negative charge is expelled outward and the kinetic energy
  of the electrons is lowered relative to that   with a non-expanded orbital.}
\end{figure}

One of several\cite{holeelec} ways to incorporate this physics in the Hubbard Hamiltonian  is by the substitution\cite{pincus,color,dynhub}
\beq
Un_{i\uparrow}n_{i\downarrow}\rightarrow (U+\alpha q_i)n_{i\uparrow}n_{i\downarrow}
\eeq
where $\alpha$ is a coupling constant (assumed positive) and $q_i$ a local boson degree of freedom  describing the orbital relaxation,
with equilibrium position at $q_i=0$ if zero or one electrons are present: upon
double occupancy of the orbital at site $i$, $q_i$ will change from zero to a negative value to reduce the electronic on-site repulsion, to an
extent determined by the boson dynamics. If we describe the dynamics of
this boson by a simple harmonic oscillator\cite{pincus}
\beq
H_i=\frac{p_i^2}{2m}+\frac{1}{2} K q_i^2+(U+\alpha q_i)n_{i\uparrow}n_{i\downarrow}
\eeq
the on-site repulsion is reduced from $U$ to $U_{eff}=U-\alpha^2/(2K)$ when $q_i$ takes the value $q_i=-\alpha/K$ corresponding to minimum energy when the site is
doubly occupied, versus $q_i=0$ when the site is unoccupied or singly occupied. The conventional Hubbard model does not allow the orbital to relax, in other words it 
corresponds to the limit $K\rightarrow \infty$ of an infinitely stiff orbital.

Superficially the site Hamiltonian Eq. (3) may look like the  conventional Holstein model giving rise to small polarons but it is in fact very different: the boson degree of freedom $q_i$ couples
to the $double$ occupancy of the site ($n_{i\uparrow}n_{i\downarrow}$)  rather than to the single occupancy ($n_{i\uparrow}+n_{i\downarrow})$.
Such a coupling, in a context where $q_i$ represented a phonon degree of freedom, was first introduced by Pincus\cite{pincus} but not often
considered thereafter.

The Hamiltonian Eq. (3) is intrinsically  electron-hole $asymmetric$\cite{dynhub}: the importance of this physics increases as the filling of the electronic energy band increases, as is
simply seen by taking the mean field expectation value of Eq. (3). In addition, the importance of this physics increases when the ionic charge is small\cite{dynhub}, since in that
case the orbital expansion is larger (for example, the orbital expansion is larger for $H^-$ than for $He$), corresponding to a smaller
stiffness parameter $K$ in Eq. (3).  These two facts imply that the importance of this physics
increases the more {\it negative charge} the system has. Notably, we will find in this paper that systems described by this Hamiltonian 
have a strong
tendency to {\it expel negative charge}, reflecting the radial expulsion of negative charge that already exists at the atomic level.  
Similarly, we will find  for the system as a whole described by a dynamic
Hubbard model  that negative charge
expulsion is associated with lowering of kinetic energy, just like in the atom. Note however that our site Hamiltonian Eq. (3), describing orbital expansion when
the value of $q_i$ is negative, does $not$ have a term explicitly describing the kinetic energy lowering of the atomic electron when $q_i$ adopts a negative value.
 
The potential (Coulomb) energy of a system of charges is minimized when the charge distribution is uniform. A non-uniform charge distribution gives rise to
electrostatic fields and an associated potential energy cost. It will be favored if this cost is compensated by a kinetic energy gain, i.e. lowering of kinetic energy.
In dynamic Hubbard models kinetic energy plays a key role, and   we will find in this paper  that dynamic Hubbard models are prone to develop charge inhomogeneity, and in extreme cases
charge separation, where  kinetic energy lowering overcompensates for the potential energy cost. Many materials of recent interest,
including high $T_c$ cuprates and manganites, exhibit charge inhomogeneity\cite{bianconi, stripes,dagotto,patches,patches2}, suggesting that dynamic Hubbard models may be
useful to describe them. In addition, the tendency to charge inhomogeneity will influence the charge distribution around grain boundaries, as we discuss in this paper.

One may ask whether  dynamic Hubbard models are really fundamentally different from many other models proposed and extensively studied in
the literature such as the conventional Hubbard model, the periodic Anderson model, the Falicov-Kimball model, the Holstein model, 
the Holstein-Hubbard model, the Fr\"{o}hlich Hamiltonian, standard polaron models, standard electron-boson models, the
t-J model, etc, so as to merit a new name and new studies. We believe most definitely yes. These conventional models are usually assumed to be
electron-hole symmetric, or at least electron-hole asymmetry does not play a key role, and the interesting physics in these models is usually driven by electronic
correlation and potential energy rather than by kinetic energy. Instead,   the essential physics of dynamic Hubbard models is
{\it electron-hole asymmetry}, and the   physics is {\it kinetic energy driven}. On the other hand some of these conventional models mentioned above,
when suitably modified, have properties in common with dynamic Hubbard models. In particular,  electron-hole $asymmetric$ polaron models\cite{asympol} arising from coupling to either 
electronic or spin or phononic degrees of freedom  in ways that break electron-hole symmetry have
much in common with the model discussed here.

Finally, it is certainly possible that some the physics of dynamic Hubbard models is also contained in some models studied in the past under a different name. 
For example, Robaszkiewicz et al\cite{rob}  studied a generalized periodic Anderson model with a wide and a narrow band, with $10$ different pieces to their initial Hamiltonian (Eq. (2.1)). After a
Lang-Firsov transformation leading to a small polaron representation they  
end up  with a Hamiltonian (Eq. (2.10)) that has $18$ different terms (Eq. 2.11), one of which bears some resemblance to a term in our Hamiltonian
(which has 3 terms rather than 18). As a consequence, the focus of that paper is very different from the physics discussed here.

\section{dynamic Hubbard models}

We can describe the physics depicted in Fig. 1 by a two-orbital tight binding model (for the unexpanded and expanded orbital)\cite{hole2,multi2}, or with a background 
spin\cite{hole1,color} or harmonic oscillator\cite{phonon1,undr} degree of freedom that is coupled to the electronic double occupancy, as in Eq. (2). 
We expect the physics to be similar for all these cases. Assuming the latter, the   site Hamiltonian is given by Eq. (3),
and the Hamiltonian can be written as
\beqn
H&=&-\sum_{i,j,\sigma}[t_{ij}c_{i\sigma}^\dagger c_{j\sigma}+h.c.]+ \sum_i \hbar \omega_0 a_i^\dagger a_i\nonumber \\
&+&\sum_i [U+g\hbar \omega_0(a_i^\dagger + a_i)]n_{i\uparrow}n_{i\downarrow}
\eeqn
with frequency $\omega_0=\sqrt{K/m}$ and $g=\alpha/(2K\hbar\omega_0)^{1/2}$ the dimensionless coupling constant. Estimates for the values of these parameters were discussed in
ref. \cite{dynhub}. In particular, for $1s$ orbitals $g^2\hbar \omega_0\sim 4.1 eV$. Quite generally we expect $g$ to increase proportionally to $1/Z$ and $\omega_0$ to increase
proportionally to $Z^2$, where $Z$ is the charge of the ion when
the orbital under consideration is empty\cite{dynhub}. However, in an even more realistic model $\omega_0$ should also   change with different electronic occupation. That issue
is  beyond the scope of this paper.

Using a generalized Lang-Firsov transformation\cite{mahan,undr} the electron creation operator $c_{i\sigma}^\dagger$ is written in terms of new quasiparticle operators
$\tilde{c}_{i\sigma}^\dagger$ as
\beqn
c_{i\sigma}^\dagger&=&e^{g(a_i^\dagger-a_i)\tilde{n}_{i,-\sigma}}\tilde{c}_{i\sigma}^\dagger=[1+(e^{-g^2/2}-1)\tilde{n}_{i,-\sigma}]\tilde{c}_{i\sigma}^\dagger \nonumber \\
&+&\tilde{n}_{i,-\sigma} \times (incoherent \;  part)
\eeqn
where the incoherent part describes the processes where the boson goes into an excited state when the electron is created at the site. 
For large $\omega_0$ those terms become small and can be neglected, and even for not so large $\omega_0$ we have found from numerical studies that
their effect does not change the low energy physics qualitatively\cite{dynhub12,dynfrank}. Hence we will ignore those terms in what follows, which amounts to keeping only ground state to ground state
transitions of the boson field.

The electron creation operator is then given by
\bmath
\beq
c_{i\sigma}^\dagger=[1+(S-1)\tilde{n}_{i,-\sigma} ]\tilde{c}_{i\sigma}^\dagger
\eeq
\beq
S=e^{-g^2/2}
\eeq
and the quasiparticle weight for electronic band filling $n$ ($n$ electrons per site) is
\beq
z(n)=(1+(S-1)\frac{n}{2})^2
\eeq
\emath
so that it decreases monotonically from $1$ when the band is almost empty to $S^2<1$ when the band is almost full. 
The single particle Green's function and associated spectral function is renormalized by the multiplicative factors on the quasiparticle operators given
 in Eq. (6a))\cite{undr}, which on the average amounts to multiplication of the spectral function
by the quasiparticle weight Eq. (6c). 
This will cause a reduction in the photoemission spectral weight at low energies from what would naively follow from the low energy effective Hamiltonian,
an effect extensively discussed in Ref. \cite{undr} and recently rediscovered in \cite{dyn13}. A corresponding reduction occurs in the
two-particle Green's function and associated low frequency optical properties\cite{undr,dynhub12}.

According to Eq. (6) 
$<c_{i\sigma}^\dagger c_{i\sigma}>=z(n) <\tilde{c}_{i\sigma}^\dagger \tilde{c}_{i\sigma}>$,
which appears to indicate that quasiparticles carry a different charge than real particles. This is however not the case, as can be seen by using Eq. (5) instead of
Eq. (6) to evaluate $<\tilde{c}_{i\sigma}^\dagger \tilde{c}_{i\sigma}>$. The incoherent part accounts for the difference and in fact the quasiparticle carries an unrenormalized unit charge
equal to that of the real particle, just as in usual Landau
theory\cite{landau}. Therefore, we can obtain the real charge distribution in the system by computing  the site occupations of the quasiparticles.

The low energy effective Hamiltonian is then
\bmath
\beq
H=-\sum_{ij\sigma}  t_{ij}^\sigma [\tilde{c}_{i\sigma}^\dagger \tilde{c}_{j\sigma}+h.c.]+U_{eff}\sum_i \tilde{n}_{i\uparrow}\tilde{n}_{i\downarrow}
\eeq
\beq
t_{ij}^\sigma=[1+(S-1)\tilde{n}_{i,-\sigma} ][1+(S-1)\tilde{n}_{j,-\sigma} ] t_{ij}
\eeq
\emath
and $U_{eff}=U-\hbar \omega_0 g^2$. Thus, the hopping amplitude for an electron between sites $i$ and $j$ is given by $t_{ij}$, $St_{ij}$ and $S^2t_{ij}$ depending on whether there
are $0$, $1$ or $2$ other electrons of opposite spin at the two sites involved in the hopping process.

The physics of these models is determined by the magnitude of the  parameter $S$, which can be understood as the overlap matrix element between the
expanded and unexpanded orbital in Fig. 1. It  depends crucially on the net ionic charge $Z$, defined as the
ionic charge when the orbital in question is unoccupied\cite{dynhub}. In Fig. 1, $Z=1$ if the 
states depicted correspond to the hydrogen ions $H^+$, $H$ and $H^-$ and  $Z=2$ if 
they correspond to $He^{++}$, $He^+$ and $He$.  In a lattice of $O^=$ anions, as in the $Cu-O$ planes of high $T_c$ cuprates, 
the states under consideration are $O$, $O^-$ and $O^=$ and $Z=0$,
and in the $B^-$ planes of $MgB_2$, $Z=1$. The effects under consideration here become larger when $S$ is small, hence when $Z$ is small.
An approximate calculation of $S$ as a function of $Z$ is given in \cite{dynhub}.

We now perform a particle-hole transformation since we will be interested in the regime of low hole concentration. 
We assume for simplicity that the Hamiltonian is defined on a hypercubic lattice with only nearest neighbor hopping, so that the particle-hole transformation leaves the
first term in the Hamiltonian Eq. (7a) unchanged by a suitable transformation of the phases. The hole creation operator is  given by, instead of Eq. (6a)
\bmath
\beq
c_{i\sigma}^\dagger=[S+(1-S)\tilde{n}_{i,-\sigma} ]\tilde{c}_{i\sigma}^\dagger
\eeq
where $\tilde{n}_{i,\sigma}$ is now the hole site occupation, and the hole quasiparticle weight increases with hole occupation $n$ as
\beq
z_h(n)=S^2(1+(\frac{1}{S}-1)\frac{n}{2})^2
\eeq
\emath

For simplicity of notation we   denote the hole creation operators again by $c_{i\sigma}^\dagger$,
the hole site occupation by $n_{i\sigma}$  and the 
effective on-site repulsion between holes of opposite spin $U_{eff}$ (the same as between electrons)  by $U$ to simplify the notation. The Hamiltonian for holes is then
\bmath
\beq
H=-\sum_{ij\sigma}  t_{ij}^\sigma [ {c}_{i\sigma}^\dagger  {c}_{j\sigma}+h.c.]+U \sum_i  {n}_{i\uparrow} {n}_{i\downarrow}
\eeq
\beq
t_{ij}^\sigma=t_{ij}^h[1+(\frac{1}{S}-1)) n_{i,-\sigma} ][1+(\frac{1}{S}-1){n}_{j,-\sigma} ] t_{ij}
\eeq
\emath
with $t_{ij}^h=S^2t_{ij}$ the hopping amplitude for a single hole when there are no other holes in the two sites involved in the hopping process. The hole hopping amplitude
and the effective bandwidth  increase
as the hole occupation increases, and so does the quasiparticle (quasihole) weight Eq. (8b).

Assuming there is only nearest neighbor hopping $t_{ij}=t$,      the nearest neighbor hopping amplitude 
resulting from Eq. (9b) is
\bmath
\beq
t_{ij}^\sigma=t_h+\Delta t(n_{i,-\sigma}+n_{j,-\sigma})+\Delta t_2 n_{i,-\sigma}n_{j,-\sigma}
\eeq
with
\beq
t_h=tS^2
\eeq
\beq
\Delta t=tS(1-S)
\eeq
\beq
\Delta t_2=t(1-S)^2=(\Delta t)^2/t_h   .
\eeq
\emath
The non-linear term with coefficient $\Delta t_2$ is expected to have a small effect when the band is close to full (with electrons)  and is often neglected.
Without that term, the model is also called the generalized Hubbard model or Hubbard model with correlated hopping\cite{kiv,camp}.
The effective hopping amplitude for average site occupation $n$ is, from Eq. (10a)
\beq
t(n)=t_h+n\Delta t+\frac{n^2}{4}\Delta t_2
\eeq
so that a key consequence of integrating out the higher energy degrees of freedom is to renormalize the hopping amplitude and hence the 
bandwidth and the effective mass (inverse of hopping amplitude), as  recently rediscovered in a different context\cite{dyn13}.

\section{generalized dynamic Hubbard models}
More generally, one can assume that the boson degree of freedom will couple not only to the double orbital occupancy 
but also to the singly occupied orbital\cite{undr}. The site Hamiltonian is then
\beq
H_i=\hbar \omega_0 a_i^\dagger a_i + \hbar \omega_0(a_i^\dagger+a_i) [g n_{i\uparrow}n_{i\downarrow}+g_0(n_{i\uparrow}+n_{i\downarrow})]
\eeq
At the atomic level, the coupling $g_0$ will appear when considering an orbital for atoms other than hydrogenic ones, and represents the modification of the states of electrons in 
other orbitals in the atom when the first electron is created in the orbital under consideration. We expect this effect to be much smaller than the modification of the state of the
electron residing in the same orbital, the physics described by $g$, hence
$g_0<<g$, particularly when the ionic charge $Z$ is small. The formal development for the site Hamiltonian given by Eq. (12) is very similar to the one discussed in the previous section and
is given in Ref. \cite{undr}. In particular, Eq. (6) becomes
\bmath
\beq
c_{i\sigma}^\dagger=[T+(S-T)\tilde{n}_{i,-\sigma} ]\tilde{c}_{i\sigma}^\dagger
\eeq
\beq
T=e^{-g_0^2/2}
\eeq
\beq
S=e^{-(g+g_0)^2/2}
\eeq
\beq
z(n)=(T+(S-T)\frac{n}{2})^2  .
\eeq
\emath
and of course $S<T$ always \cite{color} since $g>0$.
The hopping amplitudes given in the previous section are similarly modified by replacing $1$ by $T$ in various places, as discussed in Ref. \cite{undr}. $g_0$ gives a renormalization of the
quasiparticle mass, bandwidth and quasiparticle weight that is independent of band filling, and $g$ gives a band-filling dependent contribution.

Recently, a single band model with site Hamiltonian of the form Eq. (12) with $g=0$ and $g_0\neq 0$ was considered\cite{dyn13} to describe the effect of higher energy
electronic excitations on the low energy electronic physics within dynamical mean field theory\cite{kotliar}.
In our view this is an unphysical limit since we expect $g>>g_0$ quite generally. Some of the effects discussed in refs. \cite{dynhub,undr,dynhub12,dynfrank} were rediscovered
in that work\cite{dyn13}.

\section{electronic versus bosonic dynamic Hubbard models}
  \begin{figure}
\resizebox{8.5cm}{!}{\includegraphics[width=7cm]{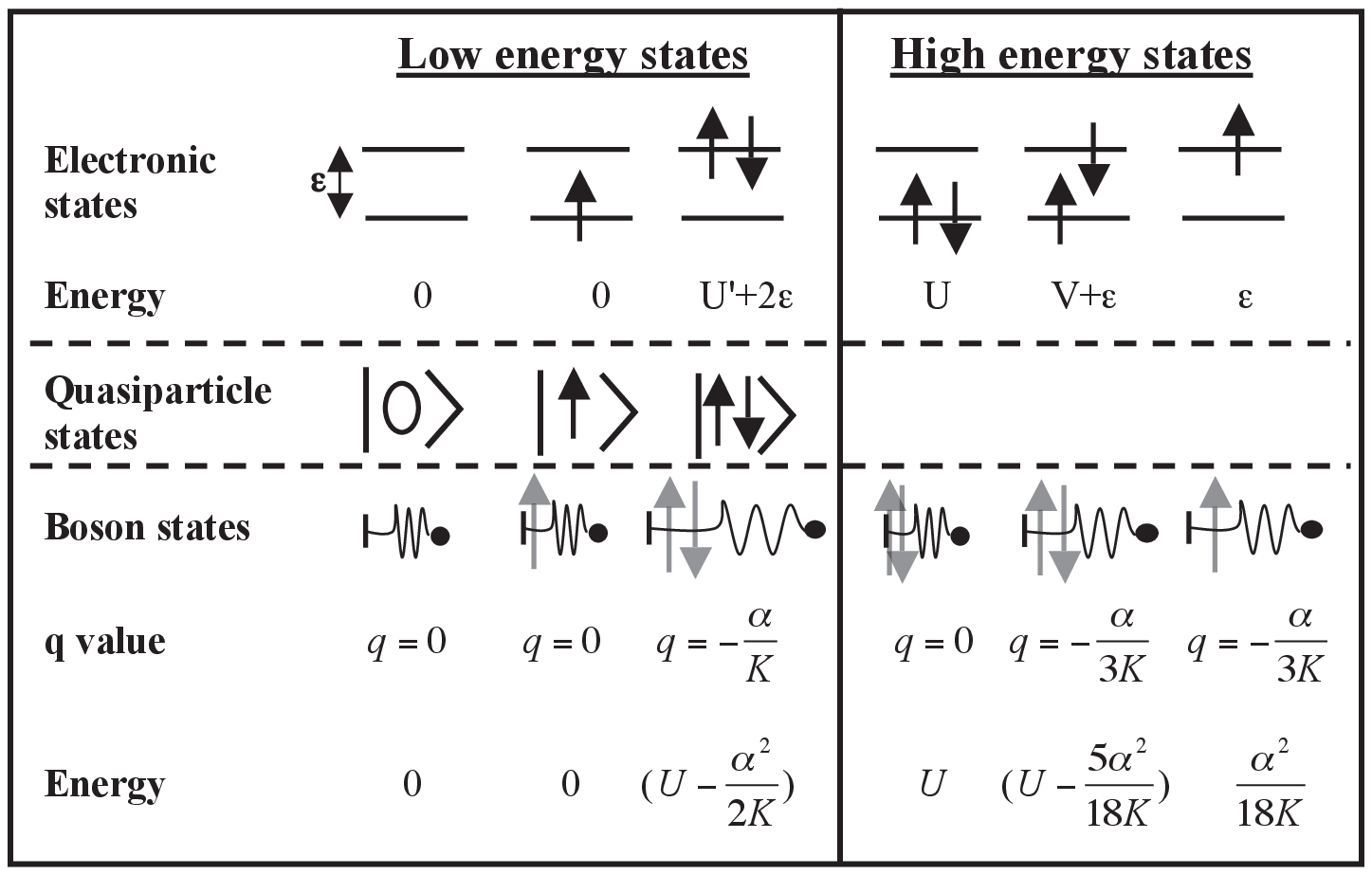}}
  \caption{Correspondence between states in the electronic dynamic Hubbard model introduced in ref. \cite{hole2} (upper part of the figure) and the bosonic
  dynamic Hubbard model discussed here (lower part of the figure). Only a few representative high energy states are shown.
  }
\end{figure} 

Here we discuss briefly the relationship between the dynamic Hubbard model with site Hamiltonian given by Eq. (3) and the electronic model with two orbitals per site
introduced in ref. \cite{hole2} and discussed further in ref. \cite{multi2}, with site Hamiltonian
\beq
H=Un_\uparrow n_\downarrow +U'n'_\uparrow n'_\downarrow +Vnn'+\epsilon n'-t'(c_\sigma^\dagger c'_\sigma+h.c.) \nonumber
\eeq
The unprimed and primed operators describe electrons in the lower and upper atomic orbitals, with single particle energy difference $\epsilon$. These orbitals
represent the unexpanded and expanded orbitals depicted in Fig. 1.

Fig. 2 shows the correspondence between the site states of the electronic and bosonic models. Parameters in the electronic model are chosen so that when the second electron is introduced at the site, the energy of the state
with  both electrons occupying the upper electronic state ($U'+2\epsilon$) is  lower than both the energy of   the state with   both electrons  in  the lower state ($U$) and 
the state with
one   electron in each
of the site states  ($V+\epsilon$). Similarly, in the bosonic model the two electron state when the oscillator is fully relaxed ($q=-\alpha/K$) has
lower energy than the state where the oscillator is unrelaxed ($q=0$) or partially relaxed ($q=-\alpha/3K$).

The details of the high energy states in both models are different, in particular the bosonic model has an infinite number of high energy states and the electronic model
only a finite number. However, the low energy effective Hamiltonian Eq. (7) is the same for both models,   and as a consequence
the charge expulsion physics discussed here is the same for both models. Furthermore the physics of spectral weight transfer from high to low energies
(undressing)\cite{undr} is the same for both models. Therefore, we argue that the electronic two-orbital model {\it with the constraints on the interaction parameters assumed}\cite{hole2} 
and the bosonic model are essentially equivalent realizations of the physics of dynamic Hubbard models.

\section{negative charge expulsion}

We consider the Hamiltonian for holes Eq. (9), with the hopping amplitudes given by Eq. (10), which we reproduce here for convenience:
\bmath
\beq
H=-\sum_{ij\sigma}  t_{ij}^\sigma [ {c}_{i\sigma}^\dagger  {c}_{j\sigma}+h.c.]+U \sum_i  {n}_{i\uparrow} {n}_{i\downarrow}
\eeq
\beq
t_{ij}^\sigma=t_h+\Delta t(n_{i,-\sigma}+n_{j,-\sigma})+\Delta t_2 n_{i,-\sigma}n_{j,-\sigma}
\eeq
\emath
It is clear from the form of this Hamiltonian that the kinetic energy decreases when the number of holes in the band increases, since the
hopping amplitudes Eq. (14b) increase with hole occupation. 
This suggest that the system will have a tendency to  {\it expel electrons} from its interior to the surface, because the coordination of sites
in the interior is larger than of sites at the surface. In what follows we study this physics numerically.

We assume a cylindrical geometry of radius R and infinite length in the z direction. We decouple the interaction terms within a simple mean field approximation 
assuming $<n_{i\sigma}>=n_i/2$ with $n_i$ the hole occupation at site $i$, and obtain the mean field Hamiltonian
\bmath
\beq
H_{mf}=H_{mf,kin}+H_{mf,pot} + H_\mu
\eeq
\beq
H_{mf,kin}=-\sum_{<ij>,\sigma}[t_h+\Delta t n_i+\Delta t_2 \frac{n_i^2}{4}] [c_{i\sigma}^\dagger c_{j\sigma}+h.c.]
 \eeq

 \beq
H_{mf,pot}=\frac{U}{4}  \sum_i n_i^2
\eeq
\beq
H_\mu=-\sum_{<ij>} n_i[\Delta t +\frac{n_j}{2}\Delta t_2] \sum_\sigma <c_{i\sigma}^\dagger c_{j\sigma}>
\eeq
\emath
Note that the local average bond occupation modifies the local chemical potential.
Assuming a band filling of $n$ holes per site, we diagonalize the Hamiltonian Eq. (15) with initial values $n_i=n$ and fill the lowest energy levels until   the occupation $n$ is achieved.
From the Slater determinant of that state we obtain new values of $n_i$ for each site and for the local bond occupation, and iterate this procedure until self-consistently is achieved.  We can  extend this
procedure to finite temperatures, iterating to self-consistency for a given chemical potential $\mu$. 
We consider then the resulting occupation of the sites as function of the distance $r$ to the center of the cylinder. Sometimes there are non-equivalent sites at the same distance from the
axis (e.g. (5,0) and (3,4)) that yield somewhat different occupation, for those cases we show the average and standard deviation as error bars in the graphs.

Figure 3 shows a typical example of the behavior found. Here we assumed $\Delta t_2=0$, corresponding to the  simpler   Hubbard model with
correlated hopping and no six-fermion operator  term. Even for $\Delta t=0$ the hole occupation is somewhat larger in the interior than near the surface.
When the interaction $\Delta t$ is turned on, the hole occupation increases in the interior and decreases near the surface. This indicates that the system expels
electrons from the interior to the surface. The effect becomes more pronounced when $\Delta t$ is increased.

Figure 4 shows the hole site occupations as circles of diameter proportional to it, for the cases $\Delta t=0$ and $\Delta t=0.25$ of Fig. 3. 
Note that the interior hole occupation is   larger for $\Delta t=0.25$ than it is for $\Delta t=0$, while near the surface the hole
occupation  is larger for $\Delta t=0$.  Again this shows that the system with $\Delta t=0.25$ is expelling electrons from the interior
to the surface, thus depleting the hole occupation near the surface.

  \begin{figure}
\resizebox{8.5cm}{!}{\includegraphics[width=7cm]{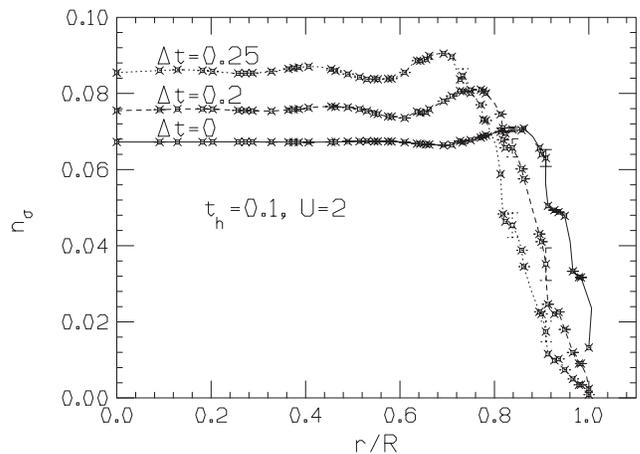}}
  \caption{Hole site occupation per spin for a cylinder of radius $R=11$ as function of $r/R$, with $r$ the distance to the center,
  for a cubic lattice of side length $1$. There are $377$ sites in a cross-sectional area ($\pi R^2=380.1$). The
  average occupation (both spins) is $n=0.126$ holes per site and the temperature is $k_BT=0.02$.}
\end{figure} 

  \begin{figure}
\resizebox{9.0cm}{!}{\includegraphics[width=7cm]{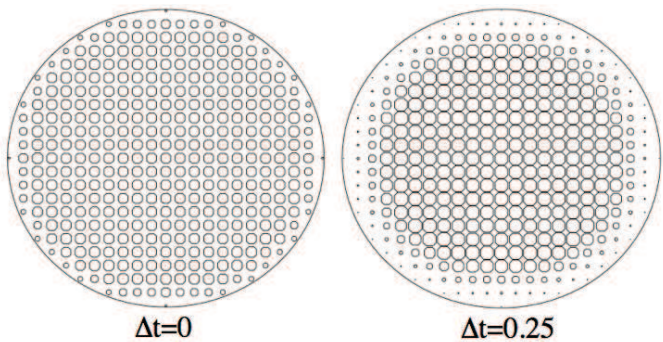}}
  \caption{The diameters of the circles are proportional to the hole occupation of the site. Note that for finite $\Delta t$ the hole occupation increases in the
  interior and is depleted near the surface. The parameters correspond to the cases shown in Fig. 3.}
\end{figure} 

These results are obtained by iteration. Fig. 5 shows the behavior of the energies as a function of iteration number for the cases $\Delta t=0$ and $\Delta t=0.25$ of Fig. 3.
The initial values correspond to a uniform hole distribution with each site having the average occupation. The evolution is non-monotonic because in the
intermediate steps the overall hole concentration increases, nevertheless it can be
seen that for the case $\Delta t=0.25$ the final kinetic energy when self-consistency is achieved is lower, and the final potential energy is higher, associated with the larger
hole concentration in the interior and the lower hole concentration near the surface shown in Fig. 4. This is of course what is expected. For the case $\Delta t=0$ instead
there is essentially no difference
in the energies between the initial uniform state and the final self-consistent state.

  \begin{figure}
\resizebox{8.5cm}{!}{\includegraphics[width=7cm]{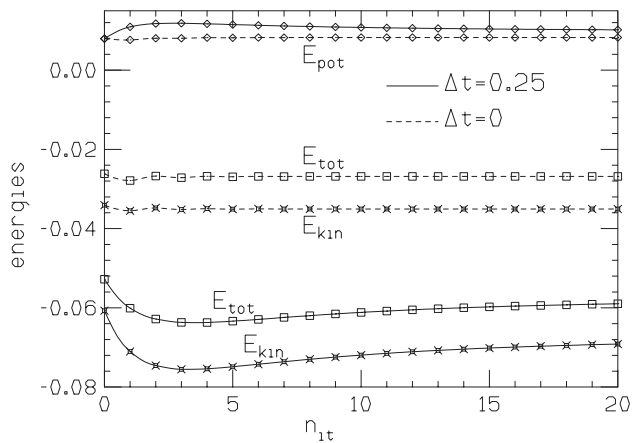}}
  \caption{Kinetic, potential and total energy per site for $\Delta t=0.25$ as function of number of iterations starting with a uniform hole distribution.}
\end{figure} 

  \begin{figure}
\resizebox{8.5cm}{!}{\includegraphics[width=7cm]{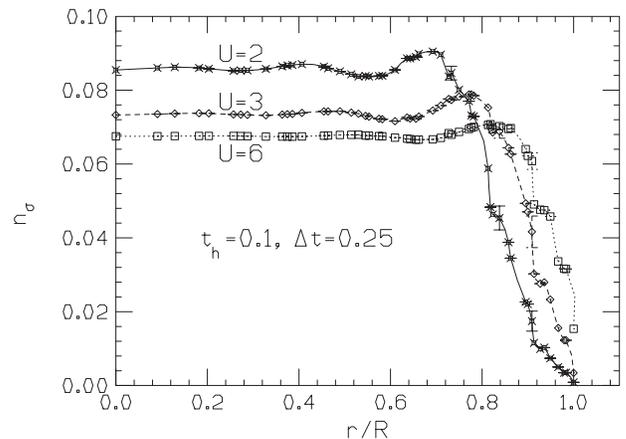}}
  \caption{Effect of Coulomb repulsion.  $t_h=0.1$,  $\Delta t=0.25$.   As the on-site repulsion
increases, the charge expulsion decreases and the occupation becomes more uniform. 
  }
\end{figure} 

 The charge expulsion caused by $\Delta t$ is counteracted by the effect of Coulomb repulsion. Figure 6 shows the effect of increasing the
 on-site repulsion for a fixed value of $\Delta t$.

The effect of the non-linear occupation-dependent hopping term $\Delta t_2$ (Eq. (10d))  is shown in Fig. 7. It increases the charge expulsion 
tendency relative to the model where this 
term is omitted (Hubbard model with correlated hopping).

As the correlated hopping amplitude $\Delta t$ increases, and even more so in the presence of $\Delta t_2$, the system appears to develop a tendency to phase
separation, where holes condense in the interior and the outer region of the cylinder becomes essentially empty of holes. This happens very rapidly as function
of the parameters for the finite system under consideration. Examples are shown in Fig. 8. We return to this point in the next section.

  \begin{figure}
\resizebox{8.5cm}{!}{\includegraphics[width=7cm]{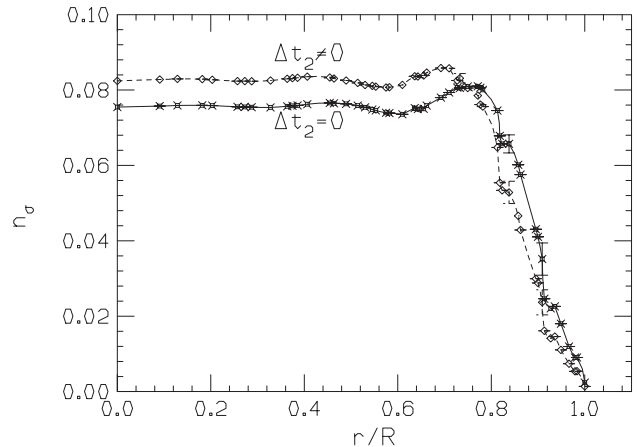}}
  \caption{Effect of non-linear occupation-dependent hopping term $\Delta t_2$, Eq. (7c)
  for the case $\Delta t=0.25$ of Fig. 3.}
\end{figure} 

  \begin{figure}
\resizebox{8.5cm}{!}{\includegraphics[width=7cm]{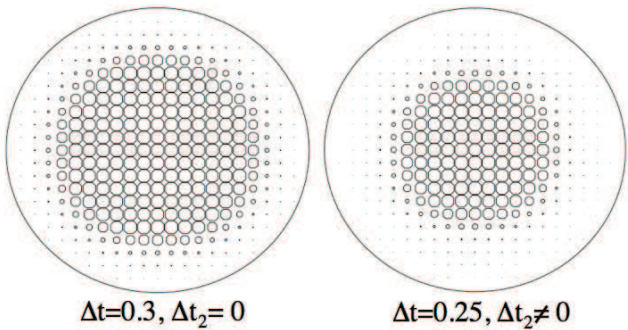}}
  \caption{As the correlated hopping terms increase, the system develops a tendency to phase separation, where essentially all the holes condense to the
  interior. Parameters are the same as in Fig. 3 except as indicated. The maximum hole occupation per spin is  0.128 and 0.214 for the left and right panel, the average hole occupation
  per spin  is 0.063.}
\end{figure} 

The charge expulsion tendency and associated effects described in this and other sections become rapidly weaker when the hole concentration increases. To illustrate this we show
in Fig. 9 the charge distribution for the same Hamiltonian parameters as Fig. 8 but average hole occupation per site $n=0.35$ instead of $n=0.126$. It can be
seen that the charge expulsion is much smaller.

  \begin{figure}
\resizebox{8.5cm}{!}{\includegraphics[width=7cm]{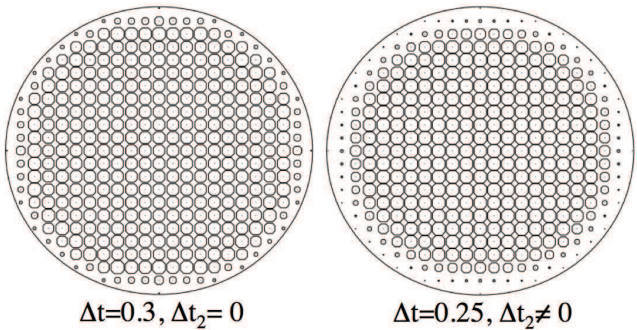}}
  \caption{Hole site occupations for the same parameters as Fig. 8 but average hole occupation per spin  0.175.
  The maximum hole occupation per spin is 0.214 for the left panel, 0.232 for the right panel. }
\end{figure}

In summary, we have seen in this section that the dynamic Hubbard model promotes expulsion of negative charge from the interior to the surface of the system when the band is almost full,
and that the charge expulsion physics is associated with kinetic energy lowering, just as in the single atom, Fig. 1.
The charge expulsion tendency is largest when the parameter $\Delta t$ is largest, which in turn corresponds to smaller $S$, the overlap of the 
atomic orbitals when one and two electrons are at the orbital. As discussed earlier, $S$ is smaller when the ionic charge $Z$ is smaller, corresponding to a 
more negatively charged ion. The fact that the effective Hamiltonian derived from this physics expels more negative charge the more negatively
charged the ion is makes of course a lot of sense and can be regarded as an internal consistency check on the validity of the model.

\section{phase separation}

The tendency to charge expulsion in the dynamic Hubbard model can also lead to a thermodynamic instability and ensuing phase separation. The question of phase separation
in these models in certain parameter ranges was recently considered by Montorsi and coworkers\cite{montorsi,montorsi2}.  

Let us consider the
correlated hopping model first for simplicity ($\Delta t_2=0$). The effective hopping amplitude for a hole is $t(n)=t_h+n\Delta t$ with $n$ the hole density per site. In the dilute
limit, the kinetic energy of a hole is $\epsilon=-zt(n)$, with $z$ the number of nearest neighbors to a site, and it is clear that when $\Delta t$ is much larger
than $t_h$ putting for example all holes into half the system and thus emptying the other half will double $n$ and hence decrease the kinetic energy per hole as well as the
total energy if it is dominated by kinetic energy. This tendency to phase separation will be countered both by Pauli exclusion and by Coulomb repulsion.

Consider a flat density of states for simplicity. The effective bandwidth increases linearly with $n$
\beq
D(n)=D_h+Kn
\eeq
with $D=2zt$, $D_h=2zt_h$, $K=2z\Delta t$. The density of states per site per spin is given by $g(\epsilon)=1/D$ and the ground state kinetic energy by
\beq
E_{kin}=\int_{-D/2}^\mu  \epsilon g(\epsilon)d\epsilon=\frac{D}{4}(n^2-2n)
\eeq
with $\mu=(n-1)D(n)/2$ the chemical potential for $n$ holes per site. Adding the on-site repulsion in a mean field approximation yields
\beq
E_o(n)=\frac{D_h+nK}{4}(n^2-2n)+\frac{U}{4}n^2
\eeq
and the system will be unstable towards phase separation into hole-rich and hole-poor regions when the condition
\beq
\frac{\partial^2E_0}{\partial n^2}=\frac{U+D_h}{2}+K(\frac{3}{2}n-1)<0 
\eeq
is satisfied, hence
\beq
K>\frac{U+D_h}{2(1-\frac{3}{2}n)}
\eeq
or equivalently
\beq
\Delta t>\frac{t_h+U/(2z)}{2(1-\frac{3}{2}n)}     .
\eeq
For the parameters used in the previous section, e.g. $t_h=0.1$, $n=0.126$, $U=2$ and $z=4$ appropriate to two dimensions Eq. (21) yields
$\Delta t > 0.216$. The tendency to phase separation becomes even stronger  when the nonlinear term $\Delta t_2$ is included. 
After some simple algebra Eq. (21) is modified to
the condition
\beq
\Delta t>\frac{t_h+U/(2z)}{2(1-\frac{3}{2}n)} -    \frac{3 \Delta t^2 n(1-n)/(4t_h)}{2(1-\frac{3}{2}n)}   
\eeq
which for the parameters given above yields $\Delta t > 0.182$. These estimates are consistent with the numerical results shown 
in the previous section. Note that as $n$ increases larger $\Delta t$ is needed for phase separation.

Note that the instability criterion Eq. (19) appears to be different from the usual criterion 
\beq
\frac{\partial \mu}{\partial n}<0
\eeq
if $\mu$ is given by the expression given right after Eq. (17). The reason is that the $\mu$ in Eq. (17) is not the true chemical potential but an effective one.
The true chemical potential is modified by contributions from both the Hubbard repulsion and the density dependent hopping terms. For the case with $\Delta t_2=0$ it is 
given by
\bmath
\beq
\mu=\mu_{eff}+\frac{Un}{2}-\frac{K}{2}<c_{i\sigma}^\dagger c_{j\sigma}+h.c.>
\eeq
\beq
\mu_{eff}=\frac{D_h+nK}{2}(n-1)
\eeq
\beq
<c_{i\sigma}^\dagger c_{j\sigma}+h.c.>=n(1-\frac{n}{2})
\eeq
\emath
where the expectation value for the bond charge Eq. (24c) follows from Eq. (17). Hence we obtain from Eq. (24)
\beq
\frac{\partial \mu}{\partial n}=\frac{U+D_h}{2}-K(1-\frac{3}{2}n)
\eeq
in agreement with Eq. (19). The instability criteria Eq. (23) or Eq. (19) with the free energy replacing $E_0$, are also valid at finite temperature of course.

In a real material in the normal state phase separation will not occur because of the effect of longer-range Coulomb interactions not included
in our model Hamiltonian. However this physics will clearly
favor charge inhomogeneity, i.e. hole-rich regions that benefit from the lowering of kinetic energy acquired by increasing the hole concentration and thereby
broadening the (local) bandwidth, and hole-poor regions where the kinetic energy cost is mitigated by narrowing of the local bandwidth. 
The shape of these regions will depend on the particular details of the system and merits further investigation. A possible 
geometry for the hole-rich and hole-poor regions could be one-dimensional, i.e. stripes\cite{stripes}. 
Other geometries like patches are also possible\cite{patches,patches2}. Such charge inhomogeneities are commonly seen in high $T_c$ superconductors\cite{bianconi}
where the physics discussed here should be dominant.

\section{charge inhomogeneity}

High $T_c$ cuprates show a high tendency to charge inhomogeneity\cite{dagotto,patches,patches2,bianconi}. We suggest that a dynamic Hubbard model may be relevant to describe
this physics: because kinetic energy dominates the physics of the dynamic Hubbard model, the system
will develop charge inhomogeneity at a cost in potential energy if it can thereby lower its kinetic energy more, unlike models where the
dominant physics is potential (correlation) energy like the conventional Hubbard model.

We assume there are impurities in the system that change the local potential at some sites, and compare the effect of such perturbations for the dynamic and
conventional Hubbard models. As an example we take parameters $t_h=0.1$, $U=2$ and consider site impurity potentials of magnitude
$\pm 0.2$ at several sites as indicated in the caption of Fig. 8. For the dynamic Hubbard model we take $\Delta t=0.2$, $\Delta t_2=0.4$, corresponding to
$S=0.333$. 

Figure 10 shows the effect of these impurities on the charge occupation for the conventional and dynamic models. In the conventional Hubbard model the occupation
 changes at the site of the impurity potential and only very slightly at neighboring sites. In the dynamic Hubbard model
the local occupation change at the impurity
site itself is much larger than in the conventional model, and in addition,   the occupations
change at many other sites in the vicinity of the impurities, as seen in the lower panel of Figure 10. Figure 11 shows the real space distribution
of these changes.

 \begin{figure}
\resizebox{8.5cm}{!}{\includegraphics[width=7cm]{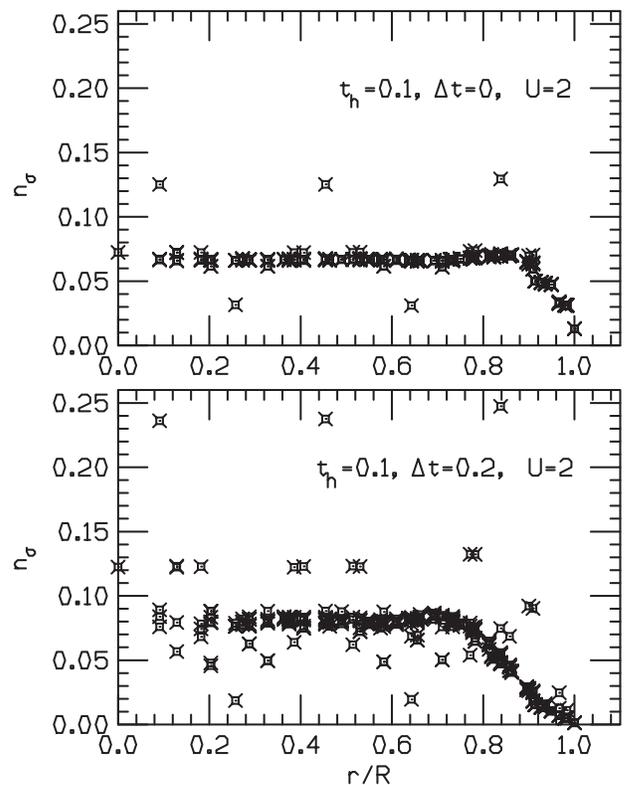}}
  \caption{Hole site occupation per spin in a system of radius $R=11$ with 5 impurities at positions
  (-1,0), (2,2), (3,-4), (-5, -5), (-6, 7) with potential strength   -0.2, +0.2, -0.2, +0.2, -0.2 respectively. Note the much larger variation in densities
  generated in the dynamic Hubbard model (lower panel, $\Delta t_2\neq 0$) than in the conventional Hubbard model (upper panel).
  Average hole occupation per site is $n=0.126$.}
\end{figure}

  \begin{figure}
\resizebox{8.5cm}{!}{\includegraphics[width=7cm]{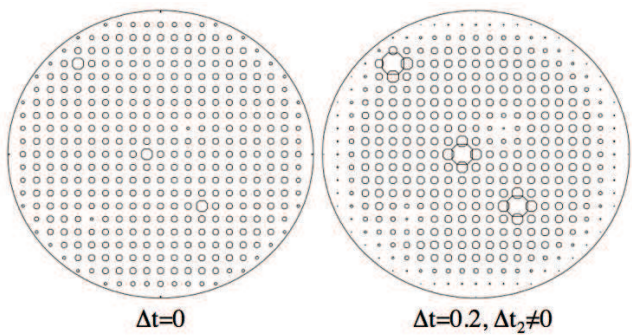}}
  \caption{The site occupations for the case of Fig. 10, with the diameters of the circles proportional to the hole occupation of the sites.
  Note the 5 impurity sites at positions listed in the caption of Fig. 10 (three with negative potential, hence higher hole concentration) and two with positive potential, hence lower hole concentration.
  Note that for $\Delta t=0$ only the occupation at the impurity site changes appreciably, while for $\Delta t \neq 0$ an impurity potential of the
  same strength causes a much larger change of the occupation at the impurity site and   occupation   change also at the nearest
  and next nearest neighbor sites.}
\end{figure} 

\begin{figure}
\resizebox{8.5cm}{!}{\includegraphics[width=7cm]{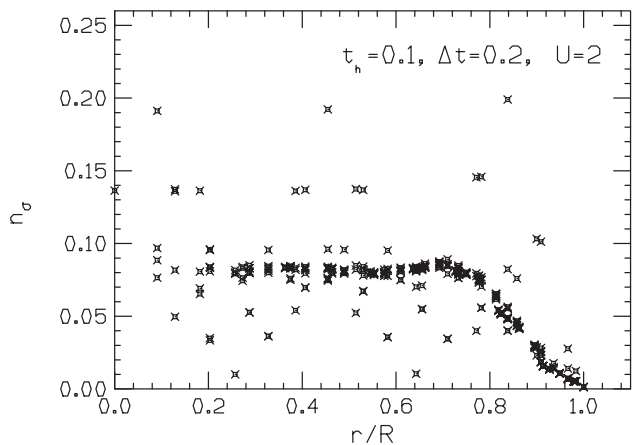}}
  \caption{Hole site occupation per spin in a system of radius $R=11$ with 5 impurities at positions
  (-1,0), (2,2), (3,-4), (-5, -5), (-6, 7) with $S$ factor 0.5, 0.2, 0.5, 0.2, 0.5 respectively. All other sites have $S=0.333$. 
  $n=0.126$.  Note that the variations in density occur for even
  more sites than when the local potential is changed, Fig. 10 lower panel, even though the change in occupation at the impurity site itself is
  smaller.}
\end{figure} 

  \begin{figure}
\resizebox{8.5cm}{!}{\includegraphics[width=7cm]{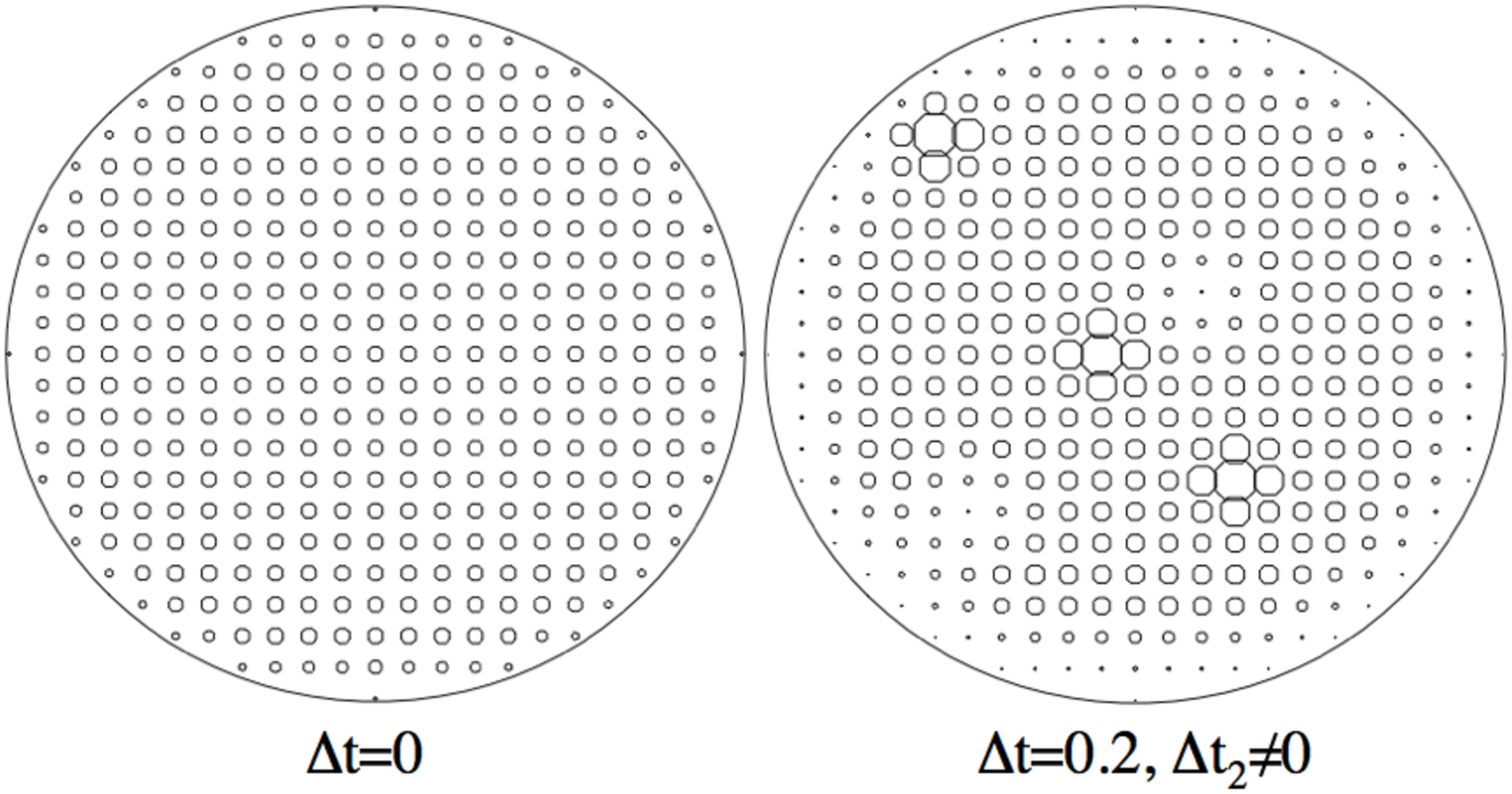}}
  \caption{The site occupations for the case of Fig. 11, with the diameters of the circles proportional to the hole occupation of the sites.
Same 5 impurity sites at positions listed in the caption of Fig. 10 (three with larger $S$,   hence higher hole concentration) and two with 
smaller $S$,  hence lower hole concentration.
  Note that the range of sites where the occupation changes is even larger than in Fig. 11, extending to third nearest neighbors. The left panel 
  shows for comparison a uniform system, corresponding to a conventional Hubbard model that does not take into account the deformation
  of the orbital, hence $S=1$ for all sites.}
\end{figure}

Similarly we can consider impurities where the atomic value of the deformation parameter $S$ is different than in the bulk. This would occur for example
by substituting an ion by another ion with different ionic charge, hence different orbital rigidity. For example, 
substituting $O^{--}$ by $F^-$ would make the orbital more rigid and increase $S$ at this site, while substituting $O^{--}$ by $N^{---}$ would make the
orbital more floppy and decrease $S$. Figure 12 shows 5 impurities at the same locations as in Figure 11, with values $S=0.5$ and $0.2$ 
at the impurity sites instead of the
bulk value $S=1/3$. The larger (smaller) $S$ will increase (decrease) the occupation. Compared to the case of Fig. 10, it can be seen that the change
in occupation at the impurity site itself is somewhat smaller for these parameters but the changes are larger at neighboring sites and extend 
to sites farther away. Similarly as in Fig. 11 we show the real space changes in Figure 13, compared to the conventional Hubbard model where
no change at all would occur since it is not sensitive to the rigidity of the orbital.

The changes in site energies in the example of Figs. 10 and 11 could occur both if there are substitutional impurities in the conducting plane with different
on-site energies, or if there are impurities off the plane that change the local potential in the plane. In contrast, the example of Figs. 12 and 13 would be
appropriate to describe only impurities in the plane itself where the electrons conduct.

The reason for this large sensitivity to local perturbations can be understood from the form of the hopping amplitude Eq. (14b). A change in the local
occupation will also change the hopping amplitude of a hole between that site and neighboring sites, which in turn will change the occupation of 
neighboring sites, and so on. Similarly a change in the deformation parameter $S$ at a site will affect the hopping amplitudes between that site
and its nearest neighbors, hence the occupation of the site and its neighbors, etc. In that way a local perturbation in the dynamic Hubbard model
gets amplified and expanded to its neighboring region, and it is easy to understand how a non-perfect crystal will easily develop patches of
charge inhomogeneity in the presence of small perturbations. These inhomogeneities cost potential (electrostatic) energy,
but are advantageous in kinetic energy. The conventional Hubbard model does not exhibit this physics.

\section{grain boundaries in high $T_c$ cuprates}

In this section we argue that dynamic Hubbard models may   be relevant to the understanding
of properties of grain boundaries in high $T_c$ cuprates\cite{bab2,hil,mann} and other materials.

Babcock et al\cite{bab} report results of EELS experiments indicating severe hole depletion in YBCO near grain boundaries
for large angle grain boundaries, with the hole deficient region extending up to $100\AA$ or more into the crystal. In contrast, small angle grain boundaries show
significantly less hole depletion. It is reasonable to assume that for larger angle grain boundaries there is weaker coupling betwen the grains, and this is confirmed experimentally
by measurement of the Josephson critical current across the grain boundary\cite{bab}. Furthermore, Babcock et al found that the structural changes associated
with the grain boundaries (structural perturbations and cation nonstochiometry) extended only about $5\AA$ from the grain boundary into the crystal, and that the hole
depletion is not associated with particular specimen preparation procedures such as time duration of oxygen annealing and storage conditions. The fact that the hole depletion
region extends over a much wider region that associated with structural changes  suggests an intrinsic purely electronic origin for the effect.
Other workers have found similar results, including Browning et al\cite{brow} and Schneider et al\cite{schn}. 

 Furthermore and consistent with this picture, it has been found that substituting $Y$ by $Ca$ near grain 
boundaries is an efficient way to prevent the hole depletion phenomenon\cite{cadoping,cadoping2,cadoping3}, since $Ca^{++}$ ions will donate fewer electrons to the $CuO$ planes than the $Y^{+++}$ ions they substitute, thus increasing the hole concentration and thereby improving the transport properties across the grain boundary.

Previous theoretical explanations of these effects have implicitly or explicitly assumed that  grain boundaries in high $T_c$ cuprates have an intrinsic {\it positive charge} that leads to band bending and consequently a flow of conduction
electrons to the vicinity of the grain boundaries that causes hole depletion\cite{bb0,bandbending,bandbending2}. However, these explanations are directly contradicted 
by experiments that  measure   the electric potential at the grain boundary by electron-beam holography\cite{holography}. 
These experiments  
show  unequivocally that the electrostatic potential at the grain boundary is $negative$ with respect to the interior\cite{negpot1,negpot2,negpot3}. 

To make sense of the electron holography results, Mannhart suggested\cite{mann} that the negative potential at the
grain boundary core may cause $overdoping$ of the
lower Hubbard band with holes, resulting in an empty band and   insulating behavior.
However, this explanation would appear to be inconsistent with the experiments of Babcock et al\cite{bab} discussed above, as well
as with the evidence that hole doping near the grain boundaries by $Ca$ substituting $Y$ improves the conduction across grain boundaries. Klie et al 
argue that their calculations\cite{klie} support the hole depletion scenario (hence positive potential at the grain boundary) and that this raises question 
about what potential is measured in the electron holography experiments. However, Koch\cite{koch} argues that ``it seems to be possible
to rule out that this observation is a measurement artifact inherent to the method of electron holography''.

\begin{figure}
\resizebox{8.5cm}{!}{\includegraphics[width=7cm]{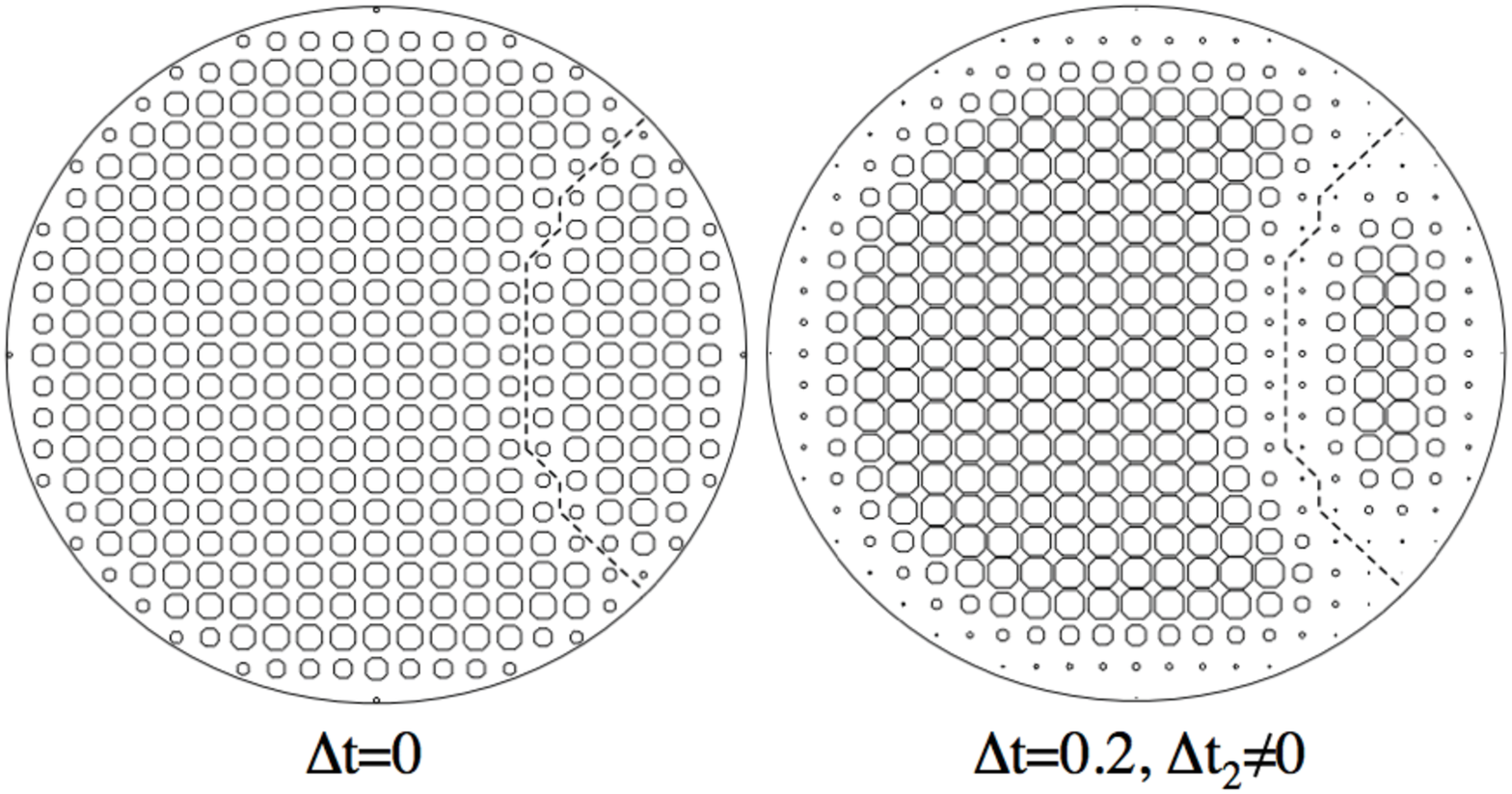}}
  \caption{Effect of a grain boundary, indicated by the dashed line for the conventional and dynamic Hubbard models with
  $t_h=0.1$, $U=2$.  We assume that the hopping amplitude is reduced by a factor $0.3$ for sites on opposite sides
  of the grain boundary. The hole occupation is depleted in the vicinity of the grain boundary in both cases, however the effect is much larger and extends over a
  wider range for the dynamic Hubbard model
  than for the conventional one.  Temperature is  $T=0.02$.}
\end{figure}

\begin{figure}
\resizebox{8.5cm}{!}{\includegraphics[width=7cm]{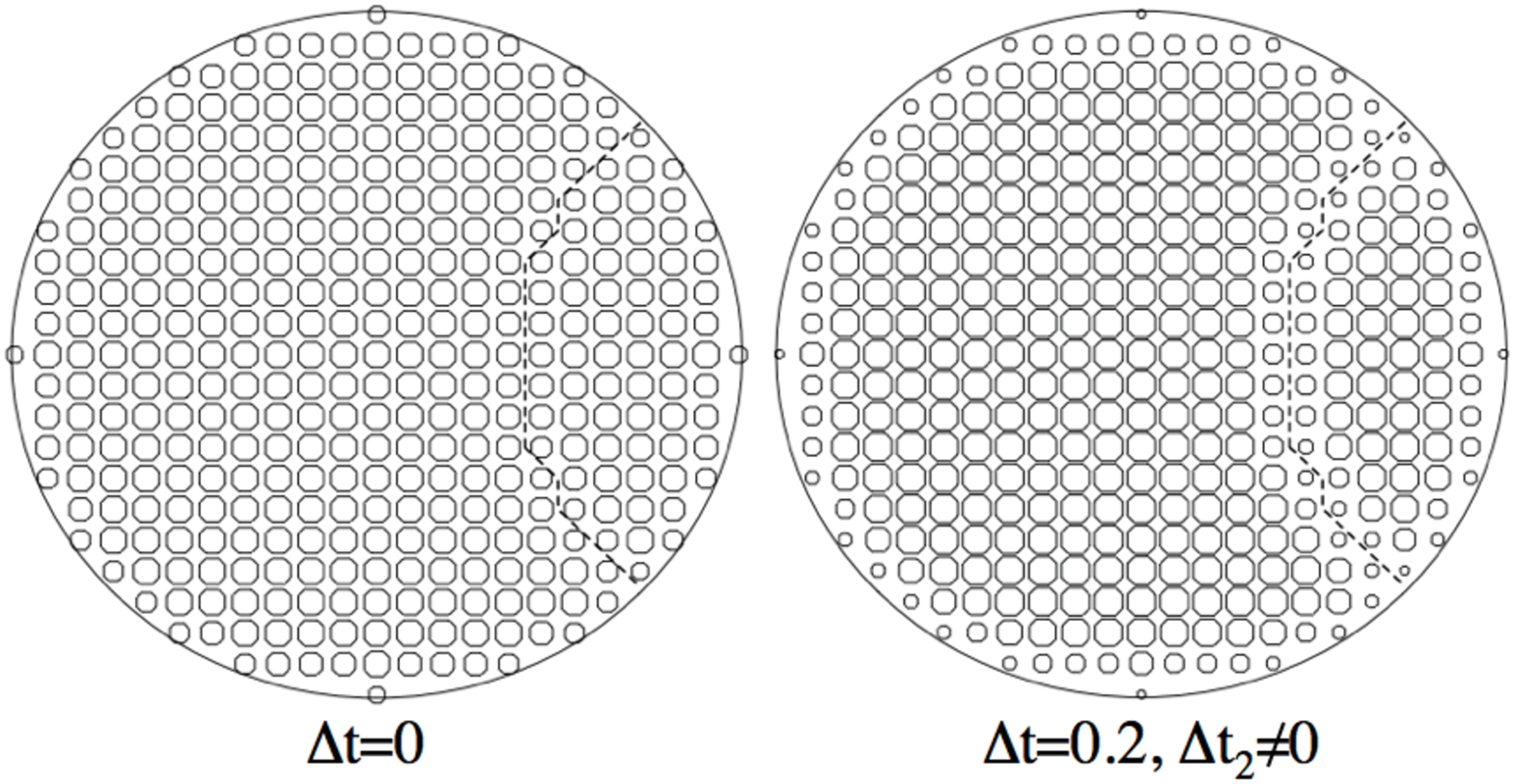}}
  \caption{Same as Fig. 14 with temperature $T=0.1$}
\end{figure}

These seemingly contradictory observations however are consistent with the physics predicted by the model
under consideration here. We model the grain boundary by assuming a smaller hopping amplitude between sites at opposite sides of the grain boundary.
A large angle grain boundary would presumably have a more reduced hopping amplitude compared to a small angle grain boundary. 
Figure 14 shows the charge distribution resulting from our model assuming a $70\%$ reduction in the hopping amplitude across the grain boundary
(denoted by a dashed line), presumably corresponding to a fairly large angle grain boundary with significant increase in the resistance. 
The hole density in the vicinity of the grain boundary is significantly reduced, driven by lowering of kinetic energy of the system. 
As a consequence, the negative charge density becomes larger near the grain boundary, and this would give rise to a {\it negative electric potential} at
the grain boundary, consistent with the electron holography experiments, and a depleted hole concentration around the grain boundary,
consistent with the EELS experiments\cite{bab}.

As a function of temperature, Schneider et al\cite{schn}  and Mennema et al\cite{menn} find that increasing the temperature leads to a rapid decrease of the grain boundary resistance. We suggests that this results from an increase in the hole concentration
near the grain boundary as the temperature is increased. Figure 15 shows that as the temperature is increased the hole density near the grain boundary increases
substantially in the dynamic Hubbard model, and as a consequence the 
conductivity of the region would increases.
It would be interesting to measure directly the dependence of hole depletion on temperature by EELS experiments.
This has not yet been done to our knowledge.

Recent measurement of properties of grain boundaries in Fe-As superconductors\cite{gbsfeas} indicate that the behavior is similar to the high $T_c$ cuprates. This 
would not be surprising if the
physics of both classes of materials  is described by dynamic Hubbard models.

\section{discussion}
Both the conventional Hubbard model and the dynamic Hubbard model are simplified descriptions of real materials, as are other models used to describe electronic
materials such as the periodic Anderson model, the Holstein-Hubbard model, the t-J model, etc. 
Whether any model contains
the physics of interest for particular real materials is in principle an open question. In this paper we have argued that the dynamic Hubbard model, which entails a 
rather straightforward and natural generalization of the conventional Hubbard model motivated by consideration of the physics of atomic orbitals in real atoms, has interesting properties in some parameter regimes 
that were not known beforehand and that
may have implications for the understanding of properties of  real
 materials.

The new physics of the dynamic Hubbard model is that it allows the electronic orbital to expand when it is doubly occupied. This expansion has
associated with it outward motion of negative charge as well as lowering of the electron's kinetic energy {\it at the atomic level}, and it is intrinsically 
$not$ ``electron-hole symmetric''. This physics is not included in the conventional Hubbard model that assumes that  the electronic orbital is infinitely rigid.
The key question is not whether this physics exists in real atoms, of course it does\cite{orbx}, the key questions are how large the effect is, what are its consequences, 
and under which conditions
and for which materials is it or is it not relevant for various properties of the materials it aims to describe.

The mathematical treatment of dynamic Hubbard models is not simple, and from the outset we restricted ourselves in this paper to the antiadiabatic limit, i.e. assuming that
the energy scale associated with the orbit expansion ($\omega_0$ in Eq. (4)) is sufficiently large than it can be assumed infinite. This brings about
the simplification that the high energy degrees of freedom can be eliminated and the Hamiltonian becomes equivalent to the low energy
effective Hamiltonian Eq. (7), a Hubbard model with correlated hoppings, linear and nonlinear terms $\Delta t$ and $\Delta t_2$. 
This low energy effective Hamiltonian, together with the quasiparticle weight renormalization described by Eq. (6), describes   properties
that we believe are relevant to real materials and are not described by the conventional Hubbard model. 

In this paper we have shown with specific quantitative examples that the dynamic Hubbard model has a strong tendency to expel negative charge from the interior of the system to the 
surface, driven by lowering of kinetic energy of the charge carriers. 
 We believe it is truly notable that this property of the model mimics the physics of the
single atom that motivated the formulation of the model, {\it even though} the kinetic energy lowering and negative charge cloud expansion of the
atomic electron is $not$ explicitly included in the site Hamiltonian Eq. (1). 
It is furthermore notable that the orbital expansion in an ion is larger when the ion is negatively charged, which in the model translates to a 
larger value of the coupling constant $g$ (smaller $S$, Eq. (6b)), larger $\Delta t/t_h$ (Eq. (10)) and consequently larger tendency
for the model to expel negative charge. 
We argue that the fact that the lattice Hamiltonian Eq. (9) describes at
a macroscopic length scale the same physics at the atomic scale that motivated the Hamiltonian is a strong indication that the Hamiltonian is relevant for the description
of reality.

The effects predicted by this Hamiltonian are largest when the coupling constant $g$ is large, or equivalently when  the overlap matrix element
$S$ is small, which corresponds to a ``soft orbital'' that would 
exist for negatively charged anions, $and$ the effects are also  largest when the band is almost full with negative electrons
(strong coupling regime). Thus, not surprisingly, more negative charge at the ion  
or/and in the band yield  larger tendency to negative charge expulsion for  the
entire system.
 We believe that the Hamiltonian is relevant to describe the physics of   materials  
including high $T_c$ cuprates, $Fe$ pnictides, $Fe$ chalcogenides, $MgB_2$ and $BiS_2$-based\cite{bis2} materials. These materials have
negatively charged ions ($O^{--}, As^{---}, S^{--}, Se^{--}, B^-$) with soft orbitals, and for most of them, including ``electron-doped'' cuprates\cite{edoped}  there is 
experimental evidence for dominant $hole$ transport in the normal state. We
  suggest that the orbital expansion and contraction of these
negative ions depending on their electronic occupation is responsible for many interesting properties of these materials and is described by the dynamic Hubbard model.

We have seen in this paper that the model leads to   charge inhomogeneity driven by lowering of kinetic energy, and in extreme cases to phase separation,
and that it leads to negatively charged grain boundaries and depletion of hole carriers in the vicinity of grain boundaries,
properties experimentally observed in many of these materials but not understood using conventional models such as ``band bending''.

Much of the physics of   dynamic Hubbard models for finite $\omega_0$ remains to be
understood. In fact, the model itself may require substantial modification to account for different values of $\omega_0$ for different electronic occupations:
the excitation spectrum of the neutral hydrogen atom, $H$,  is certainly very different from that of $H^{-}$. In connection with this and 
going beyond the antiadiabatic limit where only diagonal transitions of the auxiliary boson field are taken into account as in this paper,
it is possible that $vertical$ transitions may play a key role in understanding certain  properties of
systems described by dynamic Hubbard models

In other work we discuss the related  facts that the Hamiltonian Eq. (9) gives rise to pairing of holes and superconductivity when the electronic energy band is close to full,
driven by lowering of kinetic energy\cite{deltat}.
We have also proposed elsewhere an alternative electrodynamic description of the superconducting state arising from this model that describes expulsion of negative charge
from the interior to the surface\cite{chargeexp}. Finally, we have pointed out that   in the presence of an external  magnetic field negative charge expulsion from the interior to the surface would give rise to magnetic field
expulsion from the interior to the surface\cite{meissner,plasma}.

\acknowledgements

The author is grateful to F. Guinea for helpful discussions.


\begin{references} 
\bibitem{hub} ``The Hubbard Model: A Collection of Reprints'', ed. by A. Montorsi, World Scientific, Singapore, 1992.
\bibitem{hub2} ``The Hubbard model: its physics and mathematical physics'', ed. by D. Baeriswyl, D.K. Campbell, J.M.P. Carmelo, F. Guinea and E. Louis, 
NATO ASI Series B Vol. 343, Plenum, New York, 1995.
\bibitem{orbx} J.C. Slater, {\it Quantum Theory of Atomic Structure}, McGraw-Hill, New York, 1960.
\bibitem{hole1} J.E. Hirsch, Phys. Lett. A {\bf 134}, 451 (1989).
\bibitem{relax} A. Fortunelli and A. Painelli, Chem. Phys. Lett. {\bf 214}, 402 (1993).
\bibitem{inapp} J.E. Hirsch, Physica B {\bf 199\&200}, 366 (1994).
\bibitem{tang} J.E. Hirsch and S. Tang, Sol. St. Comm. {\bf 69}, 987 (1989).
\bibitem{hole2} J.E. Hirsch, Phys. Rev. B {\bf 43}, 11400 (1991).
\bibitem{dynhub} J.E. Hirsch, Phys.Rev. Lett.  {\bf 87}, 206402 (2001).
\bibitem{dyn3} P. Sun and G. Kotliar, Phys. Rev. B {\bf 66}, 085120 (2002).
\bibitem{dyn5}  A.S. Moskvin and Y.D. Panov, Phys. Rev. B {\bf 68}, 125109 (2003).
\bibitem{dyn7} F. Aryasetiawan et al, Phys. Rev. B {\bf 70}, 195104 (2004).
\bibitem{dyn8} L. Arrachea and A.A. Aligia, Physica C {\bf 408}, 224 (2004).
\bibitem{dyn11} K. Bouadim et al, Phys. Rev. B {\bf 77}, 014516 (2008).
\bibitem{dyn121} G. H. Bach, J.E. Hirsch  and F. Marsiglio, Phys. Rev. B {\bf 82}, 155122 (2010).
\bibitem{bach1} G. H. Bach and F. Marsiglio, Jour. Sup. Nov. Mag. {\bf 24}, 1571 (2011).
\bibitem{dyn12} G. H. Bach and F. Marsiglio, Phys. Rev. B {\bf 85}, 155134 (2012).

\bibitem{kaiser} S. Kaiser et al, arXiv:1211.7017 (2012).


\bibitem{holeelec} J.E. Hirsch, Phys.Rev. B {\bf 65}, 184502 (2002).
\bibitem{pincus} P. Pincus, Solid St. Comm. {\bf 11}, 51 (1972).
\bibitem{color} J.E. Hirsch, Physica C  {\bf 201}, 347 (1992).


\bibitem{bianconi} A. Bianconi et al, Jour. of Phys. Cond. Matt. {\bf 12}, 10655 (2000).
\bibitem{stripes} S.A. Kivelson et al, Rev. Mod. Phys. {\bf 75}, 1201 (2003).
\bibitem{dagotto} E. Dagotto, Science {\bf 309}, 257 (2005).
\bibitem{patches} S.H. Pan et al, Nature {\bf 413}, 282 (2001).
\bibitem{patches2} Y. Kohsaka et al, Nature Physics {\bf 8}, 534 (2012).



\bibitem{asympol} J.E. Hirsch, in "Polarons and Bipolarons in high Tc Superconductors and Related Materials", ed. by E.K.H. Salje, A.S. Alexandrov and W.Y. Liang, Cambridge University Press, Cambridge, 1995, p. 234.

\bibitem{rob} S. Robaszkiewicz ,  R. Micnas and J. Ranninger, Phys. Rev. B {\bf 36}, 180 (1987).



\bibitem{multi2} J.E. Hirsch,  Phys. Rev. B {\bf 67}, 035103 (2003).
 
\bibitem{phonon1} J.E. Hirsch,  Phys. Rev. B {\bf 47}, 5351.
\bibitem{undr} J.E. Hirsch, Phys. Rev. B {\bf 62}, 14487 (2000); Phys. Rev. B {\bf 62}, 14998 (2000).

\bibitem{mahan} J.D. Mahan, ``Many Particle Physics'', Third Edition, Plenum, New York, 2000.



\bibitem{dynhub12} J.E. Hirsch, Phys. Rev. B {\bf 65}, 214510 (2002); Phys. Rev. B {\bf 66}, 064507(2002).
\bibitem{dynfrank} F. Marsiglio, R. Teshima and J. E. Hirsch, Phys. Rev. B {\bf 68}, 224507 (2003).

\bibitem{dyn13} M. Casula et al, Phys. Rev. Lett. {\bf 109}, 126408 (2012).

 \bibitem{landau} D. Pines and P. Nozieres, ``The Theory of Quantum Liquids'', Vol. I, Addison-Wesley, Redwood, City 1989.    

\bibitem{kiv} S. Kivelson, W.-P. Su, J. R. Schrieffer, and A. J. Heeger,  Phys. Rev. Lett. {\bf 58}, 1899 (1987). 
\bibitem{camp} D. K. Campbell, J. Tinka Gammel, and E. Y. Loh
Phys. Rev. B {\bf 38}, 12043 (1988).


\bibitem{kotliar} G. Kotliar et al, Rev. Mod. Phys. {\bf 78}, 865 (2006).


\bibitem{montorsi} A. Montorsi, Jour. Stat. Mech. (2008)  L09001.
\bibitem{montorsi2}  A. Anfossi, C. D. E. Boschi and A. Montorsi, Phys. Rev. B {\bf 79}, 235117 (2003).



 \bibitem{bab2} S.E. Babcock  and J.L. Vargas, Ann. Rev. Mat. Sci   {\bf 25}, 193 (1995).
\bibitem{hil}H. Hilgenkamp and J. Mannhart, Rev. Mod. Phys.  {\bf 74}, 485 (2002).
\bibitem{mann} J. Mannhart, in `Thin Films and Heterostructures for Oxide Electronics', Springer, 2005, p. 251.


\bibitem{bab} S.E. Babcock et al, Physica C {\bf 227}, 183 (1994).
\bibitem{brow} N.D. Browning et al, Physica C {\bf 212}, 185 (1993).
\bibitem{schn} C.W. Schneider et al, Phys. Rev. Lett.  {\bf 92}, 257003 (2004).


\bibitem{cadoping} A. Schmehl et al, Europhys. Lett. {\bf 47}, 110 (1999).
\bibitem{cadoping2} G. Hammerl  et al, Nature {\bf 407}, 162 (2000).
\bibitem{cadoping3} G.A. Daniels, A. Gurevich and D.C. Larbalestier, Appl. Phys. Lett.  {\bf 77}, 3251 (2000).
\bibitem{bb0} J. Mannhart and H. Hilgenkamp, Supercond. Sci. Technol. {\bf 10}, 880 (1997).
\bibitem{bandbending}  A. Gurevich and E.A. Pashitskii, Phys. Rev. B {\bf 57}, 13878 (1998).
\bibitem{bandbending2} U. Schwingenschlogl and C. Schuster,  EPL {\bf 77}, 37007 (2007).

\bibitem{holography} V. Ravikumar, R.P. Rodrigues and V.P. Dravid, Phys. Rev. Lett. {\bf 75}, 4063 (1995).
\bibitem{negpot1} M.A. Schofield, L. Wu and Y. Zhu, Phys. Rev. B {\bf 67}, 224512 (2003).
\bibitem{negpot2} M.A. Schofield et al, Phys. Rev. Lett.  {\bf 92}, 195502 (2004).
\bibitem{negpot3} Ch. Joos et al,  Physica C {\bf 408-410}, 443 (2004).


\bibitem{klie} R.F. Klie et al, Nature {\bf 435}, 475 (2005).

\bibitem{koch} C.T. Koch, Int. J. Mat. Res. {\bf 101}, 1 (2010).

\bibitem{menn} S. H. Mennema et al, Phys. Rev. B {\bf 71}, 094509 (2005).

\bibitem{gbsfeas}  J.H. Durrell et al, Reports of Progress in Physics {\bf 74}, 124511 (2011) and references therein.


\bibitem{bis2} Y. Mizuguchi et al, J. Phys. Soc. Jpn. {\bf 81}, 114725 (2012).
\bibitem{edoped} 
W. Jiang et al,  Phys. Rev. Lett. {\bf 73}, 1291 (1994); 
P. Fournier et al , Phys. Rev. B {\bf 56}, 14149 (1997); Y. Dagan and R.L. Greene, Phys. Rev. B {\bf 76}, 024506 (2007).

 \bibitem{deltat}  J.E. Hirsch and F. Marsiglio, Phys. Rev. B {\bf 39}, 11515 (1989); Phys. Rev. B {\bf 62}, 15131 (2000).
 
\bibitem{chargeexp} J.E. Hirsch, Phys.Rev. B {\bf 68}, 184502 (2003); Phys.Rev. B {\bf 69}, 214515 (2004).
 
 \bibitem{meissner} J.E. Hirsch, Physica Scripta  {\bf 85}, 035704 (2012).
\bibitem{plasma} W.A. Newcomb,  Ann. Phys. {\bf 3}, 347 (1958).
 
  \end{references}
 \end{document}